# The Legacy of Hideki Yukawa, Sin-itiro Tomonaga, and Shoichi Sakata: Some Aspects from their Archives

Michiji Konuma, Masako Bando, Haruyoshi Gotoh, Hisao Hayakawa, Kohji Hirata,
Kazuyuki Ito, Kenji Ito, Kazuyuki Kanaya, Daisuke Konagaya, Taichiro Kugo,
Chusei Namba, Tadashi Nishitani, Yoshinobu Takaiwa, Masaharu Tanabashi,
Kio Tanaka, Sho Tanaka, Fumihiko Ukegawa, Tadashi Yoshikawa

Presented at the 12[th] Asia Pacific Physics Conference
held at Makuhari, Chiba, Japan on 15 July 2013

To be published in the Proceedings of the 12[th] Asia Pacific Physics Conference,
Journal of the Physical Society of Japan, Supplement in 2014

Contact: mkonuma254@m4.dion.ne.jp

Supported by JSPS KAKENHI Grant Numbers 20240073, 23240111

# The Legacy of Hideki Yukawa, Sin-itiro Tomonaga, and Shoichi Sakata: Some Aspects from their Archives

Michiji Konuma*, Masako Bando[1], Haruyoshi Gotoh[2], Hisao Hayakawa[3], Kohji Hirata[4], Kazuyuki Ito[5], Kenji Ito[4], Kazuyuki Kanaya[6], Daisuke Konagaya[7], Taichiro Kugo[8], Chusei Namba[9], Tadashi Nishitani[10], Yoshinobu Takaiwa[11], Masaharu Tanabashi[12], Kio Tanaka[13], Sho Tanaka[14], Fumihiko Ukegawa[6], Tadashi Yoshikawa[15]

*Yukawa Institute for Theoretical Physics, Kyoto University, Kyoto 606-8502, Japan and Keio University., Yokohama 223-8521, Japan*
[1] *Aichi University, Nagoya 453-8777, Japan*
[2] *Kyoto University Museum, Kyoto University, Kyoto 606-8501, Japan*
[3] *Yukawa Institute for Theoretical Physics, Kyoto University, Kyoto 606-8502, Japan*
[4] *Graduate University for Advanced Studies (Sokendai), Hayama 240-0193, Japan*
[5] *Faculty of Literature, Kyoto University, Kyoto 606-8501, Japan*
[6] *Faculty of Pure and Applied Sciences, University of Tsukuba, Tsukuba 305-8571, Japan*
[7] *Faculty of Business Administration, Ryukoku University, Kyoto 612-8577, Japan*
[8] *Faculty of Science, Kyoto Sangyo University, Kyoto 603-8555, Japan*
[9] *National Institute for Fusion Science, Toki 509-5292, Japan*
[10] *Kikuchi College of Optometry, Nagoya 461-0001, Japan*
[11] *Tsukuba University of Technology, Tsukuba 305-8521, Japan*
[12] *Kobayashi-Maskawa Institute for the Origin of Particles and the Universe, and Department of Physics, Nagoya University, Nagoya 464-8602, Japan*
[13] *Faculty of Literature, Nara Women's University, Nara 630-8506, Japan*
[14] *Kyoto University, Kyoto 606-8501, Japan*
[15] *Nagoya Women's University, Nagoya 468-8507, Japan*

E-mail: mkonuma254@m4.dion.ne.jp

Hideki Yukawa, Sin-itiro Tomonaga and Shoichi Sakata pioneered nuclear and particle physics and left enduring legacies. Their friendly collaboration and severe competition laid the foundation to bring up the active postwar generation of nuclear and particle physicists in Japan. In this presentation we illustrate milestones of nuclear and particle physics in Japan from 1930's to mid-1940's which have been clarified in Yukawa Hall Archival Library, Tomonaga Memorial Room and Sakata Memorial Archival Library.

## 1. Introduction

Hideki Yukawa (1907-1981), Sin-itiro Tomonaga (1906-1979), and Shoichi Sakata (1911-1970) pioneered the theory of nuclear and particle physics and left enduring legacies. They left huge amounts of materials on their scientific and social activities. These materials are kept and analyzed at the Yukawa Hall Archival Library (Kyoto), the

Tomonaga Memorial Room (Tsukuba) and the Sakata Memorial Archival Library (Nagoya) [1,2,3,4,5,6,7].

In this presentation we illustrate some remarkable aspects seen from these archives.

## 2. The Birth and Development of the Theory Group on Elementary Particle and Nuclear Physics in Japan

In 1932, the neutron was discovered. W. Heisenberg wrote a series of three articles on nuclear structure made of proton and neutron. Immediately Yukawa reviewed this series in details. In April 1933, Yoshio Nishina and Tomonaga reported their phenomenological analysis on scattering of neutron by proton [8] and Yukawa presented his thinking on a possible existence of electron in the nucleus at the Annual Meeting of the Physico-Mathematical Society of Japan. Since then discussions and interaction among Japanese physicists in this field have been expanded and strengthened step by step.

*2.1 Phenomenological analysis on neutron scattering by proton by Nishina and Tomonaga in 1933*

The abstract of "Scattering of neutron by proton" by Nishina and Tomonaga remains in the program of the Annual Meeting of the Physico-Mathematical Society of Japan held in Sendai, 3 April 1933 [8].

> Abstract (originally in Japanese): *We have analyzed scattering of neutron by proton using Heisenberg's theory on nuclear structure, assuming the shape of interactions between neutron and proton, and taking into account the mass defect of hydrogen 2. Our result has been compared with experimental results.*

*2.2 Tomonaga told Yukawa on details of their analysis on nuclear interaction*

Responding to Yukawa's request, Tomonaga wrote a letter of 7 pages to Yukawa explaining details of his analysis on the scattering of neutron by proton assuming various shapes of phenomenological short-range potential including present-day's Yukawa potential in May or June of 1933 [9].

In page 3 of this letter Tomonaga wrote that "in the case of $J(r) = A\, e^{-\lambda r}/r$, we gave a report in the Sendai Meeting of the Physico-Mathematical Society of Japan in April 1933" as shown in Fig.1 [10].

*2.3 Yukawa acknowledged Tomonaga in his first article on meson theory*

Yukawa's first article on meson theory was written in November 1934 (Fig.2) and published in the beginning of 1935 [11,12]. He wrote in this article that "we can calculate the mass defect of $H^2$ and the probability of scattering of a neutron by a proton ...", "These calculations were made previously, according to the theory of

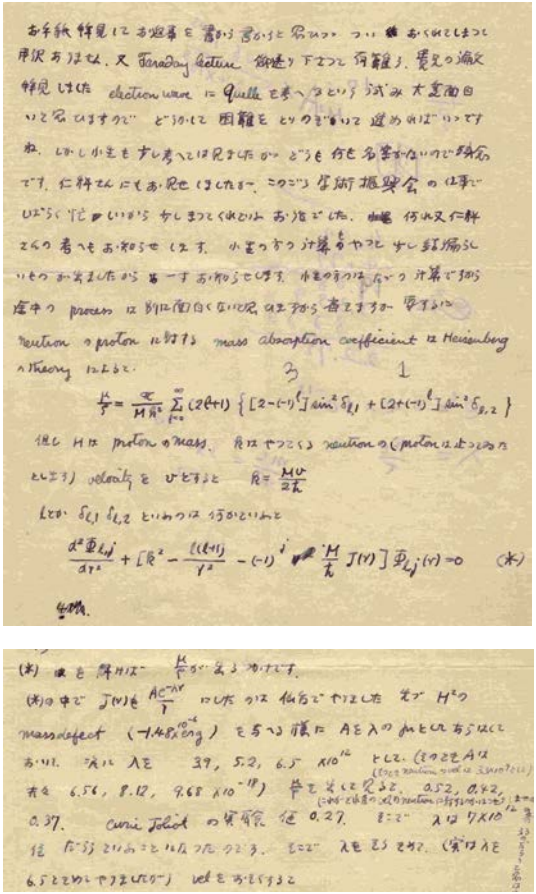

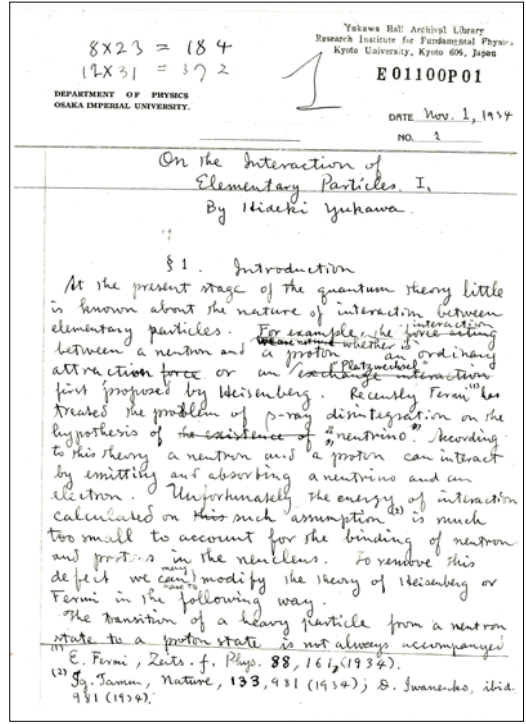

**Fig. 2.** Yukawa's manuscript for his first article on the meson theory written on 1 November 1934.

**Fig. 1.** Page 1 and a part of p.3 of the letter of Tomonaga to Yukawa in 1933. The "Yukawa potential" is explicitly mentioned in p.3.

Heisenberg, by Mr. Tomonaga, to whom the writer owes much." (p.52)

In the Yukawa Hall Archival Library, materials to trace every steps of Yukawa's research in those days are kept in a good order. The letter of Tomonaga in 1933 was preserved personally by Yukawa together with materials relating to the first article on the meson theory, before they were donated to the Archival Library.

*2.4 The cosmic-ray meson (muon) was discovered by three groups including Nishina et al.*

In 1937 three groups, S. H. Neddermeyer and C. D. Anderson [13]; Nishina, M. Takeuchi and T. Ichimiya [14, 15]; J. C. Street and E. C. Stevenson [16], discovered a new particle with mass consistent with that of Yukawa's particle in the cosmic ray by cloud-chamber experiments.

*2.5 Nishina invited Yukawa et al for an informal discussion meeting.*

**Fig. 3.** A letter of Nishina to Yukawa on 5 August 1937.

**Fig. 4.** Yukawa's Laboratory Diary on 24 April 1942.

On 5 August 1937, Nishina at RIKEN (RIkagaku KENkyusho = Institute for Physical and Chemical Research) in Tokyo sent an invitation letter (Fig.3) to Yukawa at Osaka Imperial University for an informal discussion meeting on mesons among theoretical and cosmic-ray physicists [17].

> Translation of the beginning of the letter shown in Fig.3: *How are you? Have you obtained any new achievement in your theory? May I propose a meeting to discuss in details on your theoretical results, relations between our experiments and theories and articles, in order to obtain both experimental and theoretical success in Japan?*

The proposed meeting was held at RIKEN on 19 August 1937. Yukawa and Sakata attended it from Osaka. As it was successful, similar meetings were held time by time with gradual expansion under the leadership of Nishina.

*2.6. A series of informal discussion meetings recorded by Yukawa in his laboratory diary.*

Yukawa left detailed records of research activities from 1938 to 1948 in his Laboratory Diary [18]. Seven informal discussion meetings mainly held at RIKEN were described between June 1941 and September 1943. One of them was MEISO-KAI (Wondering Meeting) on 24 April 1942 [19] (Fig.4). Yukawa reported his attempt on an extended space-time description in the quantum field theory [20] in this meeting (Fig.5). He discussed i) non-separability of cause and effect, ii) the density matrix, iii) the field equation with additional conditions, iv) Lagrangian and Hamiltonian, and v) elementary particles – numerable; space-time – point-like. He recorded even names of 21 participants and 9 questioners including Tomonaga, who got a clue to his super many-time theory by Yukawa's talk.

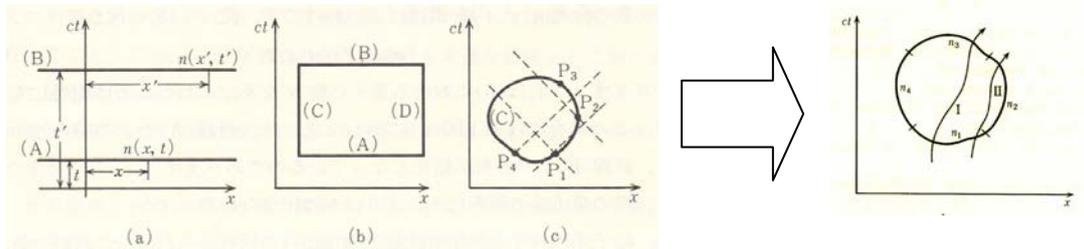

**Fig. 5.** Yukawa's attempt to an extended space-time description in the quantum field theory [20].

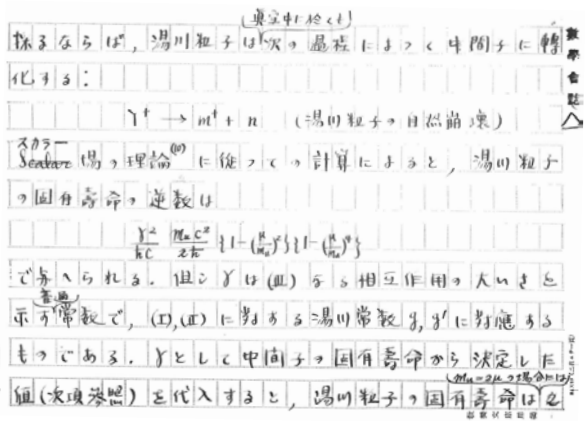

**Fig. 6.** Sakata's manuscript for the two-meson theory.

*2.7 The two-meson theory by Sakata and Yasutaka Tanikawa et al. — The prediction of $\mu$ and $\nu_\mu$ in addition to $\pi$.*

The properties of the cosmic-ray mesons became gradually clear to contradict with those required for nuclear-force mesons. Yukawa tried to change the basis of the quantum field theory. Tomonaga tried to improve the method of approximation to solve the problem. Tanikawa and Sakata discussed to change the models to get rid of the difficulty. Sakata and Takesi Inoue proposed the two-meson theory in 1942 (in Japanese, Fig.6) [21] and in 1946 (translated in English) [22], where they proposed $\mu$ (m) and $\nu_\mu$ (n) in addition to $\pi$ meson (Y). This theory was extensively discussed at the MEISO-KAI in June 1942 and finally confirmed by a cosmic ray experiment by C. Powell's group in 1947.

*2.8 Tomonaga's super many-time theory was inspired by Yukawa's attempt.*

Tomonaga was inspired by Yukawa's repeated unsuccessful attempts to make a divergence-free quantum field theory including his talk at the MEISO-KAI on 24 April 1942. Tomonaga's article on super many-time theory was published in June 1943 with the title "On a Relativistically Invariant Formulation of the Quantum Theory of

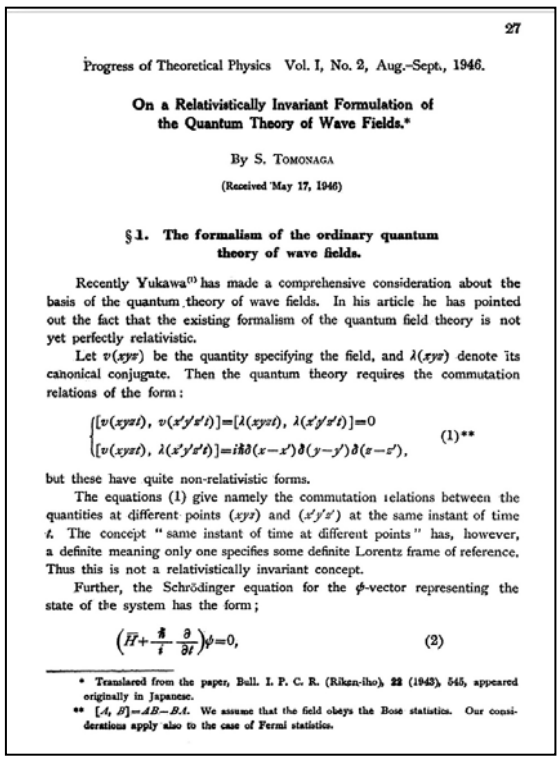

Fig. 7. The first page of Tomonaga's super many-time theory [24].

Wave Fields" in Japanese [23]. This paper was translated in English in 1946. [24] (Fig.7)

The first sentence of the paper was that "Recently Yukawa has made a comprehensive consideration about the basis of the quantum theory of wave fields. In his article he has pointed out the fact that the existing formalism of the quantum field theory is not yet perfectly relativistic".

This was the first step to the renormalization theory by Tomonaga's group.

*2.9 The Discussion Meeting on Mesons on 26 and 27 September 1943*

The largest informal discussion meeting called the Meson Symposium was held at RIKEN on 26 and 27 September 1943. Around 50 participants attended from all over Japan in spite of severe wartime. This was the last gathering in this series during the World War II. The number of speakers was 5 for 2 days including Tomonaga and Sakata. All of them prepared detailed texts, which were circulated to expected nation-wide participants in advance. Tomonaga discussed methods of approximation on nucleon-meson interactions [25] and Sakata did models of elementary particles [26]. The Proceedings of this Symposium published in 1949 [27] gave a strong impact to the next generation.

## 3. Conclusion

The creative thinking of these talented individuals and frank discussions among them in the early days gave rise to the atmosphere of friendly collaboration and severe competition, which brought up the active nuclear and particle physicists in the post WW-II generation of Japan.

## Acknowledgment

The authors acknowledge the Nishina Memorial Foundation for providing a copy of the handwritten letter of Nishina to Yukawa on 5 August 1937. They also express their thanks to the support by JSPS KAKENHI Grant Numbers 20240073, 23240111.